\documentclass[aps,prb,reprint,twocolumn,showpacs,superscriptaddress]{revtex4-2}

\usepackage{amsmath}
\usepackage{graphicx}
\usepackage{lmodern}
\usepackage{amsmath}
\usepackage{color}
\usepackage{mathtools}
\usepackage{lipsum}
\usepackage{amssymb}
\usepackage{bm}
\usepackage{braket}
\usepackage{afterpage}
\usepackage{dsfont}

\newcommand{\bk}{\bm{k}}
\newcommand{\bp}{\bm{p}}
\newcommand{\bq}{\bm{q}}

\DeclareMathOperator{\Tr}{Tr}

\newcommand{\be}{\begin{equation}}
\newcommand{\ee}{\end{equation}}

\usepackage[dvipsnames]{xcolor}
\usepackage[colorlinks=true]{hyperref}
\hypersetup{
    colorlinks=true,
    linkcolor=blue,
    citecolor=blue,
    urlcolor=blue
} 

\makeatletter
\def\maketitle{
\@author@finish
\title@column\titleblock@produce
\suppressfloats[t]}
\makeatother

\begin{document}
\title{Pair density wave order in multiband systems}
\author{Nicole S. Ticea}
\affiliation{Department of Applied Physics, Stanford University, Stanford, CA 94305, USA}

\author{S. Raghu}
\author{Yi-Ming Wu}
\affiliation{Stanford Institute for Theoretical Physics, Stanford
  University, Stanford, California 94305, USA}

\date{\today}

\begin{abstract}
	An indispensable ingredient for pair density wave (PDW) superconductivity is the presence of an attractive pairing interaction at finite momentum. Here, we show how this condition can be met with straightforward electron-density interactions in multiband systems. The electron-density interaction, when projected to the band basis, acquires form factors with nontrivial momentum dependence and thereby exhibits a potential tendency to a finite-momentum pairing instability. By applying a mean field analysis to two simple multiband models, the checkerboard lattice and three-band Hubbard model, we find that PDW order can indeed become the leading instability if a strong nearest-neighbour attraction is present. Moreover, the condition for the transition from a uniform superconductor to a PDW superconductor is shown via a simple quantum geometric argument.  
\end{abstract}
\maketitle
\section{\MakeUppercase{Introduction}} 
\label{sec:introduction}
Pair density wave (PDW) superconductors are comprised of Cooper pairs with finite center-of-mass (COM) momenta. This constitutes an exotic superconducting phase in which the order parameter acquires periodic spatial modulation and vanishes on average \cite{Agterberg}, thereby breaking translational symmetry on top of the expected $U(1)$ symmetry of a generic superconductor. Unlike uniform superconductivity, PDW pairing is not typically considered a weak coupling instability since the bare pairing susceptibility at finite momentum does not diverge logarithmically. Exceptions can arise when the Fermi surface is nested in the pairing channel \cite{Hsu2017,Wu2023,PhysRevB.107.224516,PhysRevB.108.035135,Huang2022,PhysRevLett.129.167001,PhysRevB.105.L100509}, or in two dimensional systems tuned to the vicinity of a higher order Van Hove singularity such that the bare finite-momentum pairing susceptibility is promoted to a {power-law} divergence \cite{HOVHS,PhysRevLett.131.026601}. On generic grounds, however, PDW order is viewed as an intermediate coupling phenomenon. Attempts to elucidate its microscopic mechanism have proven challenging. 

The first example of finite-momentum superconductivity was discussed in the context of superconductors exposed to strong magnetic fields, where the opposite-spin pairs inevitably possess finite total momentum due to the strong Zeeman splitting of the spin-up and spin-down Fermi surfaces \cite{FF,LO}. This is the FFLO (Fulde-Ferrell-Larkin-Ovchinnikov) phase. Modern interest in the PDW state, which is distinguished from FFLO physics by the absence of a magnetic field, was revitalized by a proposal suggesting its existence in under-doped cuprates \cite{Berg2007,Wang2015a,Wang2015b,Wang2018,PhysRevX.4.031017,PhysRevB.97.174511,PhysRevLett.99.067001,PhysRevB.78.174529,Yang2013,Shi2021}. Indeed, recent experiments in two La-based cuprates seem to confirm its existence \cite{Lacuprate}, although a satisfactory microscopic mechanism is still lacking from a theoretical point of view. Besides the cuprates, an increasing number of experimental results recently point to PDW states manifesting in a variety of materials, including heavy-fermion UTe$_2$ \cite{Gu2023,Aishwarya2023,aishwarya2023visualizing}, iron-based superconductor EuRbFe$_4$As$_4$ \cite{Zhao2023} and kagome metal CsV$_3$Sb$_5$ \cite{Chen2021,PhysRevB.108.L081117,schwemmer2023pair,wilson2023av3sb5}. From a theoretical perspective, the PDW ground state is inherently interesting on account of the fact that it can give rise to different vestigial orders, such as charge-density-wave and charge-$4e$ superconductivity through the process of partial melting \cite{Agterberg2008,Berg2009,PhysRevB.84.014513,RevModPhys.87.457,wu2023dwave}. Therefore, understanding the microscopic mechanism of PDW beyond weak coupling theory contributes to our understanding of strongly-correlated systems with competing orders more broadly. 

The difficulty of proposing a microscopic mechanism for PDW formation is due, in part, to a lack of consistent evidence from a wide range of experiments that one would ideally leverage to establish the correctness of any particular perspective. The rarity of the PDW state can be understood as a consequence of the stringent requirements for a finite-momentum pairing instability. Such an instability requires attractive BCS pairing interaction at nonzero momentum, and the interaction strength must surpass a threshold value. Although the latter condition can be satisfied in strongly-interacting systems, the first condition is met with greater difficulty. One natural solution is to consider systems with large pair-hopping interactions \cite{PhysRevB.40.6745,Wardh2017,PhysRevLett.130.026001,Setty2023}. This type of interaction depends naturally on the Cooper pair COM momentum $\bq$, and can therefore lead to a PDW instability if it becomes sufficiently attractive at some finite $\bq$. Systems of this kind can be realized in, e.g., the BEC limit of strong pairing systems \cite{CHEN20051}, the strong-coupling limit of the Holstein-Hubbard model \cite{PhysRevLett.125.167001}, the strong-coupling limit of the Kondo-Heisenberg model \cite{Berg2010,Julian,liu2023pairdensitywave}, and in twisted bilayer semiconductors if one of the two layers supplies charge-$2e$ excitons for mediating the electron-electron interaction in the other layer \cite{PhysRevB.102.235423,PhysRevLett.131.056001}.

In contrast, in systems where the electron-density interaction dominates, the Cooper channel interaction usually does not depend on the COM momentum $\bq$. However, as we will show in this paper, this is not true for multiband models. In such systems, the electron-density interaction picks up momentum-dependent form factors upon transforming from the lattice basis to the band basis. Although the bare interaction is itself independent of the COM $\bq$, the unitary transformation form factors generically depend on $\bq$, and as a result so too does the band-projected pairing interaction. 

In this paper, we supply a mean field theory of PDW superconductivity in multiband systems, with implementation on two typical lattice models: the extended Hubbard model on a checkerboard lattice and the extended three-band Hubbard model on a square lattice. By decomposing the interaction in different channels, we obtain phase diagrams illustrating the competition between various charge, spin, and pairing orders. Our work highlights the importance of the $\bm{q}$-dependent form factors in multiband systems for inducing a PDW instability;
this explains, in part, why in some recent works a PDW state is claimed to exist in multiband systems \cite{PhysRevB.103.195150,PhysRevB.81.012504,Yerin2023,Chen2023}, such as the three-band Hubbard model \cite{HCthreeband} and a spinless Honeycomb-lattice model with nearest-neighbour and next-nearest-neighbour interactions \cite{jiang2023pair}. 

In Sec. \ref{sec:method}, we begin by outlining a general prescription for analysing competing orders in multiband systems. Later, in Sec. \ref{sec:results}, we present numerical evidence for a leading PDW instability in the checkerboard and three-band Hubbard models. In Sec. \ref{sec:the_role_of_quantum_geometry} we discuss how the local curvature of the quantum metric can be tuned to favour PDW formation over uniform SC, and offer some concluding remarks in Sec.\ref{sec:conclusion}.

\section{\MakeUppercase{mean field theory of competing orders in multiband system}}
\label{sec:method}
We begin with a discussion of competing orders in multiband extended Hubbard models, with the aim of identifying a general scheme for favouring PDW formation. The Hamiltonian for a generic multiband lattice model can be written as the sum of a tight-binding part ($H_0$) and an interacting part ($H_\text{int}$), where
\begin{equation}
    \begin{aligned}
        H_0&=\sum_{ij;ab;\alpha}h_{ij;ab}^\alpha c^\dagger_{ia\alpha}c_{jb\alpha},\\
        H_\text{int}&=\sum_{i,a}U_a n_{ia\uparrow}n_{ia\downarrow}+\frac{1}{2}\sum_{\braket{ia,jb}}V_{ab}n_{ia}n_{jb}
    \end{aligned}
\end{equation}
 Here, $i,j$ denote Bravais lattice sites; $a,b$ represent sublattice degrees of freedom (assuming $s$-orbitals on each site for simplicity); and $n_{ia}$ is the density operator for sublattice $a$ at site $i$, defined as $n_{ia}\equiv\sum_\alpha n_{ia\alpha}=\sum_{\alpha}c^\dagger_{ia\alpha}c_{ia\alpha}$, with $\alpha=\uparrow,\downarrow$ representing the spin degrees of freedom.
 We introduce nearest-neighbour interaction $V$ in addition to the onsite Hubbard $U$.

For a given realization of the lattice, one can 
rewrite the Hamiltonian in the momentum basis and subsequently project into the band basis. This procedure is effected by the unitary transformation $c_{a\alpha}(\bk)=\sum_n u_{a n}(\bk)\psi_{n\alpha}(\bk)$, with the operator $\psi_{n\alpha}(\bk)$ creating a fermion with spin $\alpha$ and momentum $\bk$ in band $n$. In this new basis, the single-particle Hamiltonian is diagonal: $H_0=\sum_{n,\bk}\varepsilon_{n}(\bk)\psi_{n,\bk}^\dagger\psi_{n,\bk}$. The interaction, meanwhile, transforms as: 
\begin{align}
H_\text{int}&=\sum_{ab,\alpha\beta}\sum_{nn',mm'}\int_{\bq,\bp,\bk}V_{ab}(\bq)\label{eq:Hinformf}\\
        &~~~~~~~~~\times u^*_{na\alpha}(\bk)u^*_{mb\beta}(\bp) u_{m'b\beta}(\bp-\bq)u_{n'a\alpha}(\bk+\bq)\nonumber\\
        &~~~~~~~~~\times\psi_{n\alpha}^\dagger(\bk)\psi_{m\beta}^\dagger(\bp)\psi_{m'b\beta}(\bp-\bq)\psi_{n'a\alpha}(\bk+\bq)\nonumber
\end{align}
Through this projection, the interaction has acquired the form factors $\{u_{na}(\bk)\}$, resulting in a nontrivial momentum dependence of the effective interaction. 

We now make some simplifying assumptions. First, we assume that the bands are well-separated from each other; this allows us to focus on the sole band crossing the Fermi level. This consideration is valid in the low energy limit where the interaction strength does not exceed the band gap. Henceforth we will drop the band indices, and set $n=n'=m=m'$ in Eq.\ref{eq:Hinformf}. Furthermore, will assume that $H_0$ is spin-degenerate, so that the form factors do not depend on the spin index. This analysis can be easily generalized to systems with spin polarization.

In order to discuss the relevant symmetry-breaking orders, we need to decompose $H_\text{int}$ into charge, spin, and pairing channels. Such a decomposition is justified if we assume the transfer momentum, $\bq$, is a small deviation from the ordering vector. We first express $H_\text{int}$ in the direct and exchange channels, and then use the $SU(2)$ completeness relation to further separate the exchange channel into its charge and spin contributions. After a straightforward manipulation, we obtain
\begin{equation}
    \begin{aligned}
    H^\text{c}_\text{int}=\int_{\bq,\bp,\bk} &V_{\bq}^\text{c}(\bk,\bp)\left(\sum_{\alpha}\psi^\dagger_\alpha(\bk-\bq/2)\psi_\alpha(\bk+\bq/2)\right)\\
    &\times\left(\sum_{\beta}\psi^\dagger_{\beta}(\bp+\bq/2)\psi_{\beta}(\bp-\bq/2)\right)\label{eq:Hc}
\end{aligned}
\end{equation}
for the charge channel,
\begin{equation}
    \begin{aligned}
        H^\text{s}_\text{int}=\int_{\bq,\bp,\bk} &V_{\bq}^\text{s}(\bk,\bp)\left(\sum_{\alpha\gamma}\psi^\dagger_\alpha(\bk-\bq/2)\bm{\sigma}_{\alpha\gamma}\psi_\gamma(\bk+\bq/2)\right)\\
        &\times\left(\sum_{\beta\delta}\psi^\dagger_\beta(\bp+\bq/2)\bm{\sigma}_{\beta\delta}\psi_\delta(\bp-\bq/2)\right)\label{eq:Hs}
    \end{aligned}
\end{equation}
for the spin channel, and 
\begin{equation}
    \begin{aligned}
        H^\text{p}_\text{int}=\int_{\bq,\bp,\bk} &V_{\bq}^\text{p}(\bk,\bp)\psi^\dagger_{\uparrow}(\bk+\bq/2)\psi_{\downarrow}^\dagger(-\bk+\bq/2))\\
        &\times\psi_{\downarrow}(-\bp+\bq/2)\psi_{\uparrow}(\bp+\bq/2)\label{eq:Hp}
    \end{aligned}
\end{equation}
for the pairing channel. The corresponding interactions in the charge, spin, and pairing channels, respectively, are given by
\begin{equation}
    \begin{aligned}
        {V}_{\bq}^c(\bk,\bp)&={V}^{\text{di}}_{\bq}(\bk,\bp)-\frac{1}{2}{V}^{\text{ex}}_{\bq}(\bk,\bp)\\
        {V}_{\bq}^s(\bk,\bp)&=-\frac{1}{2}{V}^{\text{ex}}_{\bq}(\bk,\bp),
    \end{aligned}
\end{equation}
and 
\begin{equation}
    \begin{aligned}
        {V}^{\text{p}}_{\bq}(\bk,\bp) &= \sum_{ab}V_{ab}(\bk-\bp) u^*_{am}(\bk+\bq/2) u^*_{bm}(-\bk+\bq/2)\\
        &~~~~~\times u_{am}(\bp+\bq/2) u_{bm}(-\bp+\bq/2),
    \end{aligned}
\end{equation}
where the interactions in the direct and exchange channels are given by
\begin{equation}
    \begin{aligned}
        {V}_{\bq}^{\text{di}}(\bk,\bp) &= \sum_{ab}{V_{ab}(\bq)} u^*_{am}(\bk-\bq/2) u^*_{bm}(\bp+\bq/2)\\
        &~~~~~\times  u_{am}(\bk+\bq/2) u_{bm}(\bp-\bq/2),\\
        {V}_{\bq }^{\text{ex}}(\bk,\bp) &= \sum_{ab}V_{ab}(\bp-\bk)  u^*_{am}(\bk-\bq/2) u^*_{bm}(\bp+\bq/2)\\
        &~~~~~\times  u_{am}(\bp-\bq/2) u_{bm}(\bk+\bq/2).
         \label{ph_int}
    \end{aligned}
\end{equation}

Note that all the band-projected interactions, $V_{\bq}^\tau(\bk,\bp)$ (with $\tau=\text{c,s,p}$), satisfy $V_{\bq}^\tau(\bp,\bk)={V_{\bq}^\tau}^*(\bk,\bp)$; that is to say, we can view $V_{\bq}^\tau(\bk,\bp)$
as the $(\bk,\bp)$-th entry of a Hermitian matrix $V_{\bq}^\tau$. This permits us to decompose the interaction into its eigenbasis, 
\begin{equation}
    V_{\bq}^\tau(\bk,\bp)=\sum_j v_{\bq,j}^\tau(\bk)\lambda_{\bq,j}^\tau {v_{\bq,j}^{\tau}}^*(\bp),
\end{equation}
where $\lambda_{\bq,j}^\tau$ is the $j$-th real eigenvalue of $V_{\bq}^\tau$, and $v_{\bq,j}^\tau(\bk)$ is its corresponding eigenvector. After diagonalizing $V_{\bq}^\tau$ in this way, the interaction Hamiltonian can be compactly expressed as 
\begin{equation}
     H^\tau=\sum_j\int_{\bq}\lambda_{\bq,j}^\tau \left(\int_{\bk}v_{\bq,j}^\tau(\bk)(...)_{\bk,\bq}\right)\left(\int_{\bp}{v_{\bq,j}^\tau}^*(\bp)(...)_{\bp,\bq}\right)
 \end{equation} 
 Here, $(...)_{\bk,\bq}$ and $(...)_{\bp,\bq}$ represent the various fermion bilinears from Eqs.\eqref{eq:Hc}, \eqref{eq:Hs} and \eqref{eq:Hp}. In our convention, negative $\lambda_{\bq,j}^\tau<0$ signifies attraction.
 
 Since the $j$ sectors are decoupled, one can investigate symmetry breaking in different $j$-channels separately. We introduce a Hubbard-Stratonovich (HS) field $\Delta_j^\tau(\bq)$ to decouple the four-fermion interaction, and then integrate out the fermions to arrive at an effective action for the bosonic HS field. The quadratic term in the static limit is given by
\begin{equation}
    |\lambda_{\bq,j}^\tau|^{-1}(1\pm\lambda_{\bq,j}^\tau\Pi_j^\tau(\bq))|\Delta_j^\tau(\bq)|^2\label{eq:quadratic}
\end{equation}
The $+$ sign corresponds to the pairing channel, while the $-$ sign is reserved for the charge and spin channels. $\Pi_{j}^\tau(\bq)$ is the static susceptibility for the $j$-th channel. When considering charge or spin orders, this quantity is given by 
\begin{equation}
    \begin{aligned}
        \Pi_{j}&^{\tau=({c,s})}(\bq) = \int_{\bk}  v_{\bq,j}^\tau(\bk){v_{\bq,j}^\tau}^*(\bk)\\
        &\times\frac{n_F[\xi_m(\bk-\bq/2)]-n_F[\xi_m(\bk+\bq/2)]}{ \xi_m(\bk-\bq/2)-\xi_m(\bk+\bq/2)}\text{tr}\left(\Lambda^\tau\cdot\Lambda^\tau\right)
    \end{aligned}
\end{equation}
Here, $\xi_{m}(\bk)=\epsilon_m(\bk)-\mu$ is the $m$-th band dispersion measured from the chemical potential $\mu$, and $n_F(T)$ is the Fermi distribution function at temperature $T$. The $\Lambda^\tau$ matrices are $\Lambda_{\alpha\beta}^\text{c}=\delta_{\alpha\beta}$ and $\Lambda_{\alpha\beta}^\text{s}=\bm{\sigma}_{\alpha\beta}$, which means that spin summation in the charge channel yields a factor of $2$, whereas the spin channel picks up a factor of $6$ ($2$ if the spin SU$(2)$ symmetry is broken down to U$(1)$). In the pairing channel, the susceptibility is given by
\begin{equation}
    \begin{aligned}
        \Pi_{j}^\text{p}&(\bq) = \int_{\bk}  v_{\bq,j}^\text{p}(\bk){v_{\bq,j}^\text{p}}^*(\bk)\\
        &\times\frac{1-n_F[\xi_m(-\bk+\bq/2)]-n_F[\xi_m(\bk+\bq/2)]}{  \xi_m(-\bk+\bq/2)+\xi_m(\bk+\bq/2)}.
    \end{aligned}
\end{equation}
In our convention, the particle-hole bubble $\Pi_{j}^\text{c,s}(\bq)$ is dominantly negative (it reduces to $-\nu_0$ in the $\bq\to0$ limit where $\nu_0$ is the density-of-states at the Fermi level), while the particle-particle bubble $\Pi_{j}^\text{p}(\bq)$ remains positive. 

\begin{figure}[h]
\centering
\includegraphics[width=0.4\textwidth]{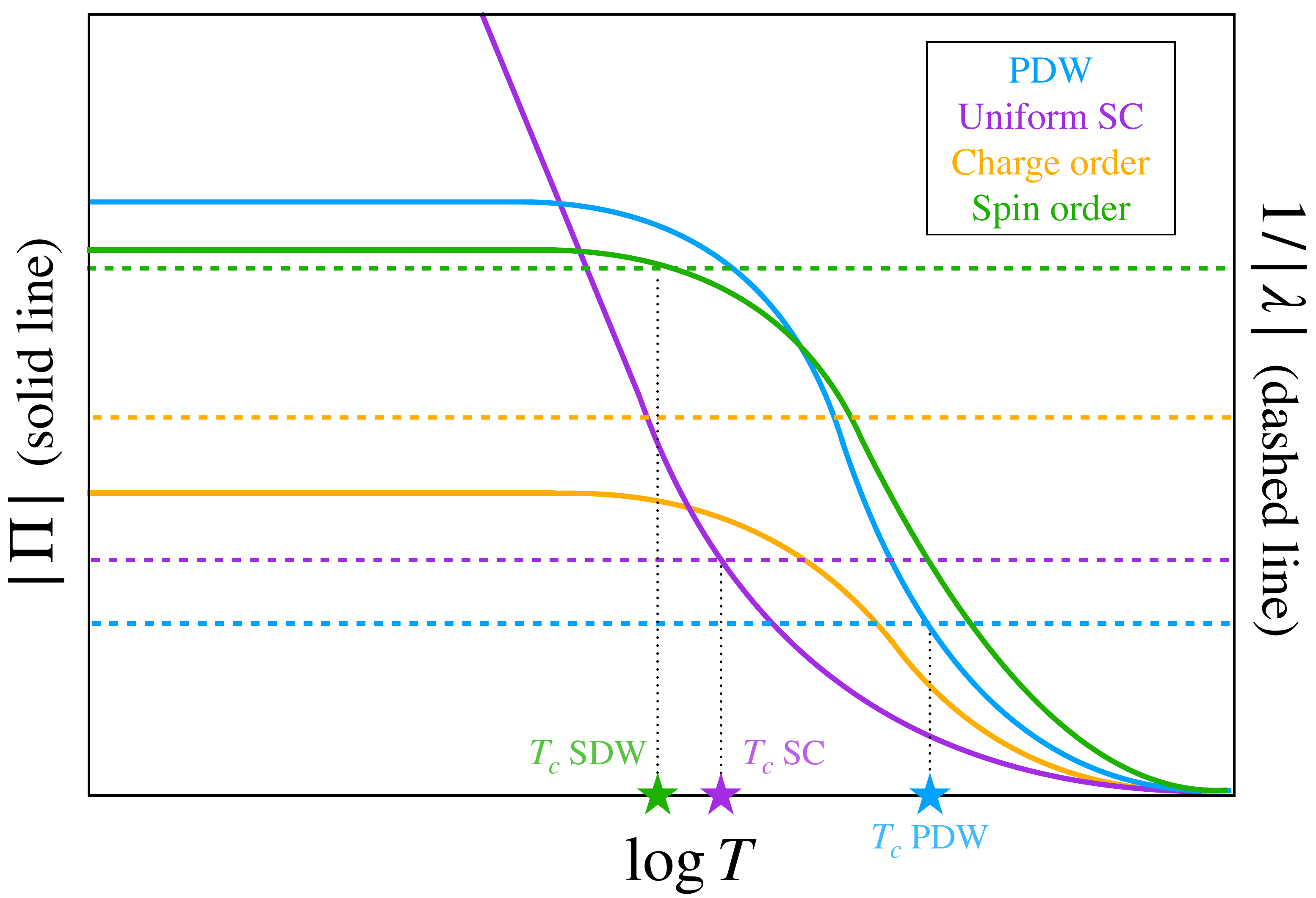}
\caption{A schematic demonstrating how contributions from the form factor may lead to a Stoner-type instability towards PDW order. Solid lines are the static susceptibilities corresponding to the PDW (blue), uniform SC (purple), spin (green), and charge (orange) channels. Dashed horizontal lines represent the inverse effective interaction, $1/|\lambda^\tau_{j,\bq}|$. In this example, the PDW has the highest $T_c$, and is therefore the leading instability in this system.}
\label{fig:stoner}
\end{figure}

One can now easily read off the conditions for superconducting or density wave ordering, whether at finite or zero COM $\bq$. The onset of a phases is determined by a transition temperature, $T_c$, at which the product $\lambda^\tau_{\bq,j}\times\Pi^\tau_{\bq,j} = -1$. While spin, charge, and superconducting orders may all co-exist, the dominant order is determined by the one with the highest $T_c$; this is the Stoner criterion. We schematize this logic in Figure \ref{fig:stoner}. 
By invoking special Fermi surface physics, one can engineer situations in which PDW order competes with uniform SC down to $T\to 0$. This is possible, for example, in the presence of higher-order Van Hove singularities, where all the bare susceptibilities diverge in a power-law manner \cite{HOVHS}. However, we stress that our result does not depend on any special behaviour of the static pair susceptibility, which as $T\to 0$ generically diverges for $\Pi_{\bq=0}^{\text{p}}$, or saturates at some finite value for $\Pi_{\bq\neq 0}^{\text{p}}$ and $\Pi_{\bq}^{\text{c,s}}$.

\begin{figure*}
    \centering
    \includegraphics[width=0.85\textwidth]{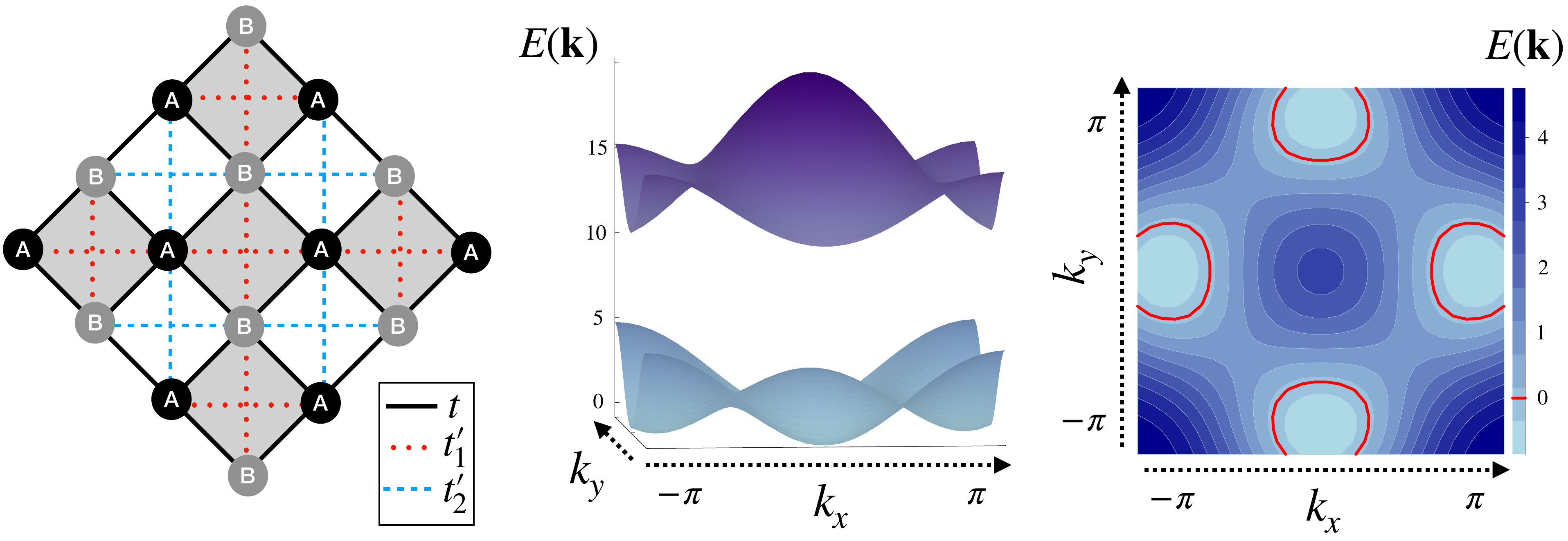}     
    \caption{(Left) Checkerboard lattice. (Middle) Band structure obtained using $t=1$, $t_1'=-0.5$, $t_2'=0.9, \phi=\pi/8$, and a $C_4$ symmetry breaking term $\Delta\sigma_z$ with $\Delta=5$. (Right) Contour plot at $1/10$ filling. The Fermi surface is identified in red.}
    \label{fig:chckb}
\end{figure*}

\section{\MakeUppercase{Results}} 
\label{sec:results}
In this section, we discuss the application of this mean field scheme to some concrete lattice examples: extended Hubbard model on a checkerboard lattice and a three-band extended Hubbard model. All hopping and interaction parameters are given in units of $t$.

\subsection{Checkerboard lattice} 
\label{sub:checkerboard_lattice}
The checkerboard lattice is a commonly studied two-band lattice model, recently attracting attenton for its tunable topological properties \cite{PhysRevLett.103.046811,kaisun2018}. Physical platforms for its realization include optical lattices \cite{wirth2010,wirth2011} and monolayer Cu$_2$N \cite{Cu2N}. Here, we use the model proposed in \cite{kaisun2018}. Each unit cell of the checkerboard lattice is composed of two sub-lattices, marked $A$ and $B$ in Fig.\ref{fig:chckb}. In momentum space, the Hamiltonian is given by $H = -\sum_{\bk} \gamma^\dagger_{\bk}\mathcal{H}(\bk)\gamma_{\bk}$, where $\gamma_{\bk}=(A_{\bk},B_{\bk})$ is a two-component spinor. $\mathcal{H}$ is a $2\times 2$ matrix:
\begin{align}
    &\mathcal{H} = \Big[(t_1'+t_2')(\cos k_x + \cos k_y)  \Big]\sigma_0 \nonumber\\
    &+ 4t\cos\phi\left(\cos\frac{k_x}{2}\cos\frac{k_y}{2} \right)\sigma_x + 4t\sin\phi\left(\sin\frac{k_x}{2}\sin\frac{k_y}{2} \right)\sigma_y \nonumber\\
    &+\Big[(t_1'-t_2')(\cos k_x - \cos k_y)+\Delta\Big]\sigma_z
\end{align}
The operator $A^\dagger_{\bk}$ ($B^\dagger_{\bk}$) creates an electron on the $A$ ($B$) sublattice. Next-nearest-neighbour hopping is parameterized by $t_1'$ on the shaded square plaquette and by $t_2'$ on the blank plaquette. The system possesses four-fold rotational symmetry, with a rotation center located at the center of the plaquettes. This $C_4$ symmetry, in combination with time-reversal symmetry (TRS), protects a quadratic band crossing point at $\bm{M}=(\pi,\pi)$.
This quadratic band touching point can be viewed as a combination of two Driac points, and thus hosts $2\pi$ Berry flux.
If, in addition, $t_1'+t_2'=0$, the system has additional particle-hole symmetry. In Fig.\ref{fig:chckb}, we plot the band structure and Fermi surface in the presence of both a $C_4$-symmetry-breaking term and a time-reversal-symmetry-breaking term, which together conspire to gap out the band structure. The resulting bands are topologically non-trivial with Chern number $\pm 1$ \cite{kaisun2018}.

The interacting part of the Hamiltonian includes the on-site Hubbard interaction $U_A$ and $U_B$, which we set equal to each other for simplicity. We also include a nearest-neighbour interaction $V_{AB}$ and next-nearest-neighbour interaction $V'_{AB}$. For practical calculations, we have set $V_{AB}=V_{BA}$ and $V'_{AB}=V'_{BA}$.

\begin{figure}
    \centering
    \includegraphics[width=0.45\textwidth]{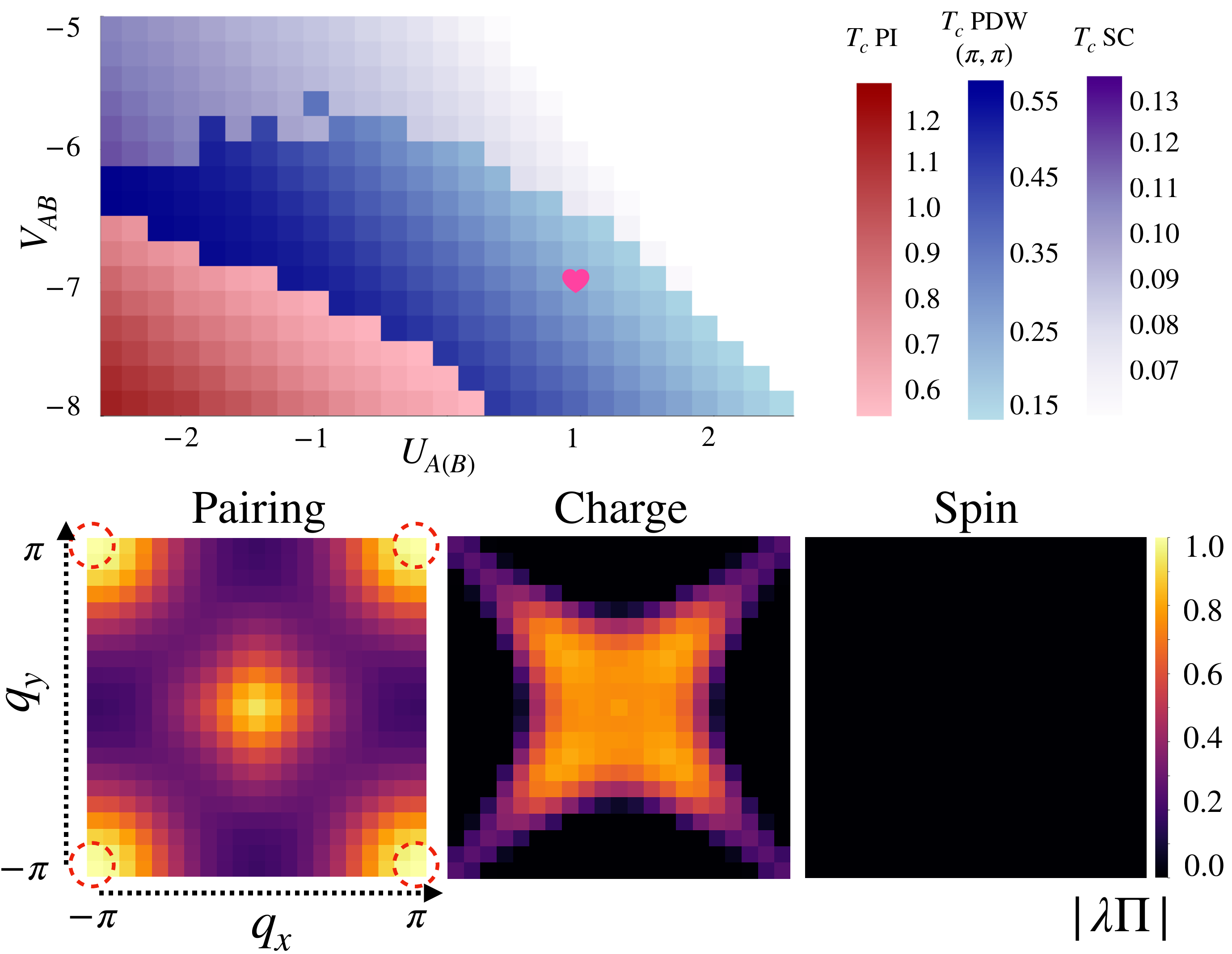}     
    \caption{(Top) Phase diagram for the parameters specified in the main text. Darker shading indicates a higher transition temperature. (Bottom) Coefficients of the pairing, charge, and spin order parameters as a function of ordering wavevector $\bq$ for $V_{AB}=-7,U_{A}=U_B=1, V'_{AA}=V'_{BB}=3.5$ at $T_c=0.27$. The location in the phase diagram corresponding to this choice of interaction strengths is marked by a pink heart. Wavevectors in the $(q_x,q_y)$ plane where the Stoner criterion is met are marked with red dashed lines. PDW = $(\pi,\pi)$ pair density wave, PI = Pomarenchuk instability, SC = uniform superconductor.}
    \label{fig:chckb_results}
\end{figure}

\begin{figure*}
    \centering
    \includegraphics[width=0.85\textwidth]{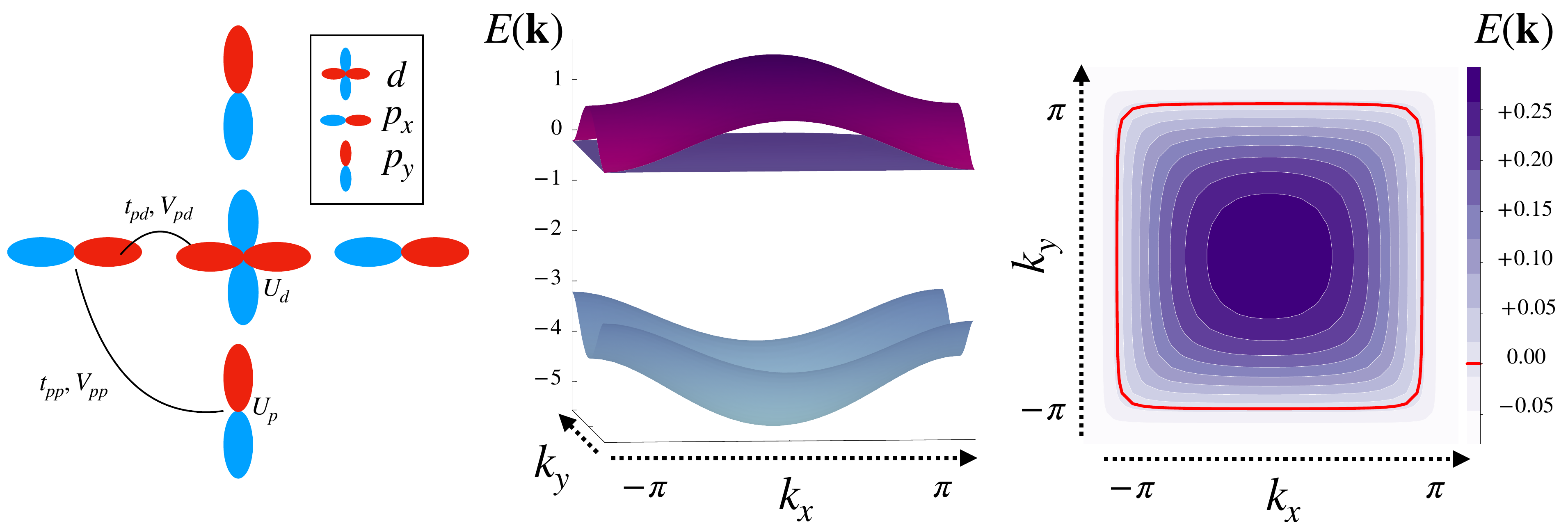}
    \caption{(Left) Microscopic lattice model of the three-band Hubbard model. (Middle) Band structure of the non-interacting part, obtained with a canonical set of parameters: $t_{pd}=1, t_{pp}=0.5$ and $\Delta_{pd}=3$. (Right) Contour plot of the second band at $1/12$ hole doping away from half-filling. The Fermi surface is identified in red.}
    \label{fig:3bH}
\end{figure*}
Results for the checkerboard lattice are summarized in Fig.\ref{fig:chckb_results}, for the same choice of parameters that were used to generate Fig.\ref{fig:chckb}. We schematize the phase diagram as a function of interaction parameters, observing the emergence of a PDW phase for attractive nearest-neighbour and both repulsive and attractive on-site interactions. Next-nearest-neighbour interactions are fixed to be repulsive; we find this choice to be necessary for stabilizing PDW formation against competing charge orders. This set of parameters can be realized in a case when strong Coulomb interaction are combined with some form of Holstein-phonon-induced attraction. Evidence of such nearest-neighbour attraction has been found in the 1D cuprates \cite{1Dcuprate}.  
 
 The PDW phase here competes with both $s$-wave uniform superconductivity and a Pomeranchuk instability (PI) in the charge channel. Also in Fig.\ref{fig:chckb_results}, we plot the coefficients of the pairing, charge, and spin order parameters as a function of ordering wavevectors $\bq$. The corresponding point in parameter space is indicated with a pink heart in the phase diagram. For this choice of interaction strengths, we observe that only the pairing channel with $\bq=(\pi,\pi)$ meets the Stoner criterion ($|\lambda^\tau_{\bq,j}\times \Pi^\tau_{\bq,j}|>1$), indicating a leading $(\pi,\pi)$ PDW instability. By examining $v^\text{p}_{\bq,0}$, the eigenvector corresponding to the most negative eigenvalue in the pairing channel at pairing vector $\bq$, we can identify the pairing symmetry for this PDW superconductor; here, we find that $v^\text{p}_{\bq=(\pi,\pi)}$ does not carry momentum dependence, so the leading instability is $s$-wave in character.

\subsection{Three-band extended Hubbard model} 
\label{sub:three_band_extended_hubbard_model}

The three-band Hubbard model has been proposed as a minimal model for capturing the essential physics of the high-$T_c$ cuprate superconductors \cite{allen1985,emery1987,rice1988}. Recent density matrix renormalization group (DMRG) calculations of the lightly-doped three-band Hubbard model on two-leg square cylinders suggest a PDW ground state \cite{HCthreeband,Jiang23b} enhanced by nearest-neighbour attractions. In this section, we ask if this three-band Hubbard model can engender PDW order within the scope of the mean field scheme described here. The model, sketched out in Fig.\ref{fig:3bH}, is
\begin{equation}
    \begin{aligned}
    &H_0 = -t_{pd}\sum_{\langle i j\rangle,\sigma}\left(d^{\dagger}_{i\sigma}p^x_{j\sigma} + d^{\dagger}_{i\sigma}p^y_{j\sigma}+h.c.\right) \\
    &- t_{pp}\sum_{\langle i j\rangle,\sigma}\left(p^{x\dagger}_{i\sigma}p^{y}_{j\sigma}+h.c.\right)+\Delta_{pd}\sum_{i}\left(p^{x\dagger}_{i\sigma}p^{x}_{\sigma}+p^{y\dagger}_{i\sigma}p^{y}_{\sigma} \right),
\end{aligned}
\end{equation}
with interactions 
\begin{equation}
    \begin{aligned}
    H_\text{int}&=U_d \sum_{i}n^{d}_{i\uparrow}n^{d}_{i\downarrow}+U_p \sum_{i}\left(n^{p_x}_{i\uparrow}n^{p_x}_{i\downarrow}+n^{p_y}_{i\uparrow}n^{p_y}_{i\downarrow}\right)\\
    &+\frac{V_{pd}}{2}\sum_{\langle ij\rangle}\left(n^d_i n^{p_x}_j +  n^d_i n^{p_y}_j \right) + \frac{V_{pp}}{2}\sum_{\langle i j\rangle}n^{p_x}_i n^{p_y}_j\\
    &+ \frac{V_{pp}'}{2}\sum_{\langle ij\rangle}\left( n^{p_x}_i n^{p_x}_j+n^{p_y}_i n^{p_y}_j\right) + \frac{V_{dd}'}{2}\sum_{\langle i j\rangle}n^{d}_i n^{d}_j
    \end{aligned}
\end{equation}

Holes on copper $d$ and oxygen $p_x$, $p_y$ orbitals are created at lattice site $i$ with spin $\sigma$ by the operators $d^\dagger_{i\sigma}, p^{x\dagger}_{i\sigma}$, and $p^{y\dagger}_{i\sigma}$, respectively; the corresponding density operators are $n_i^d\equiv \sum_{\sigma}d^{\dagger}_{i\sigma}d_{i\sigma}$ and $n_i^{p_{x(y)}}\equiv \sum_{\sigma}p^{x(y)\dagger}_{i\sigma}p^{x(y)}_{i\sigma}$. We use the notation $\langle i j\rangle$ to denote nearest-neighbour sites. The parameters $t_{pd}$ and $t_{pp}$ control the hopping between nearest-neighbour copper-oxygen and oxygen-oxygen orbitals; $\Delta_{pd}\equiv \varepsilon_p - \varepsilon_d$ is the energy difference between having a hole on the oxygen $(\varepsilon_p)$ versus copper $(\varepsilon_d)$ site; $U_p$ and $U_d$ are the on-site Coulomb repulsion strengths for oxygen and copper orbitals, respectively; and $V_{pd}$ and $V_{pp}$ are the nearest-neighbour copper-oxygen and oxygen-oxygen interactions. Next-nearest-neighbour interactions between $d$- ($V_{dd}'$) and $p$-orbitals ($V_{pp}')$ are also considered. 

We take a canonical set of parameters, identical to those selected in \cite{HCthreeband}: $t_{pd}=1, t_{pp}=0.5, U_d=8, U_p=3$, and $\Delta_{pd}=3$. The phases of the orbitals have been
fixed such that the signs of the hopping matrix elements remain
the same throughout the lattice and are positive, as in Refs.\cite{White2015,HCthreeband}. The filling is set to $1/12$ hole doping away from half-filling. We focus our analysis on the second band only, since it is flatter and therefore has a higher density of states. While the chemical potential at this doping intersects the third band as well, the transition temperatures of any competing orders are well below those originating in the second band \footnote{However, this does not discount the possibility of inter-orbital pairing involving both the second and third bands, which may potentially serve as a source of competing orders to the PDW state identified here. We defer an analysis of inter-band phenomena to future works.}.

\begin{figure}[h]
\centering
\includegraphics[width=0.47\textwidth]{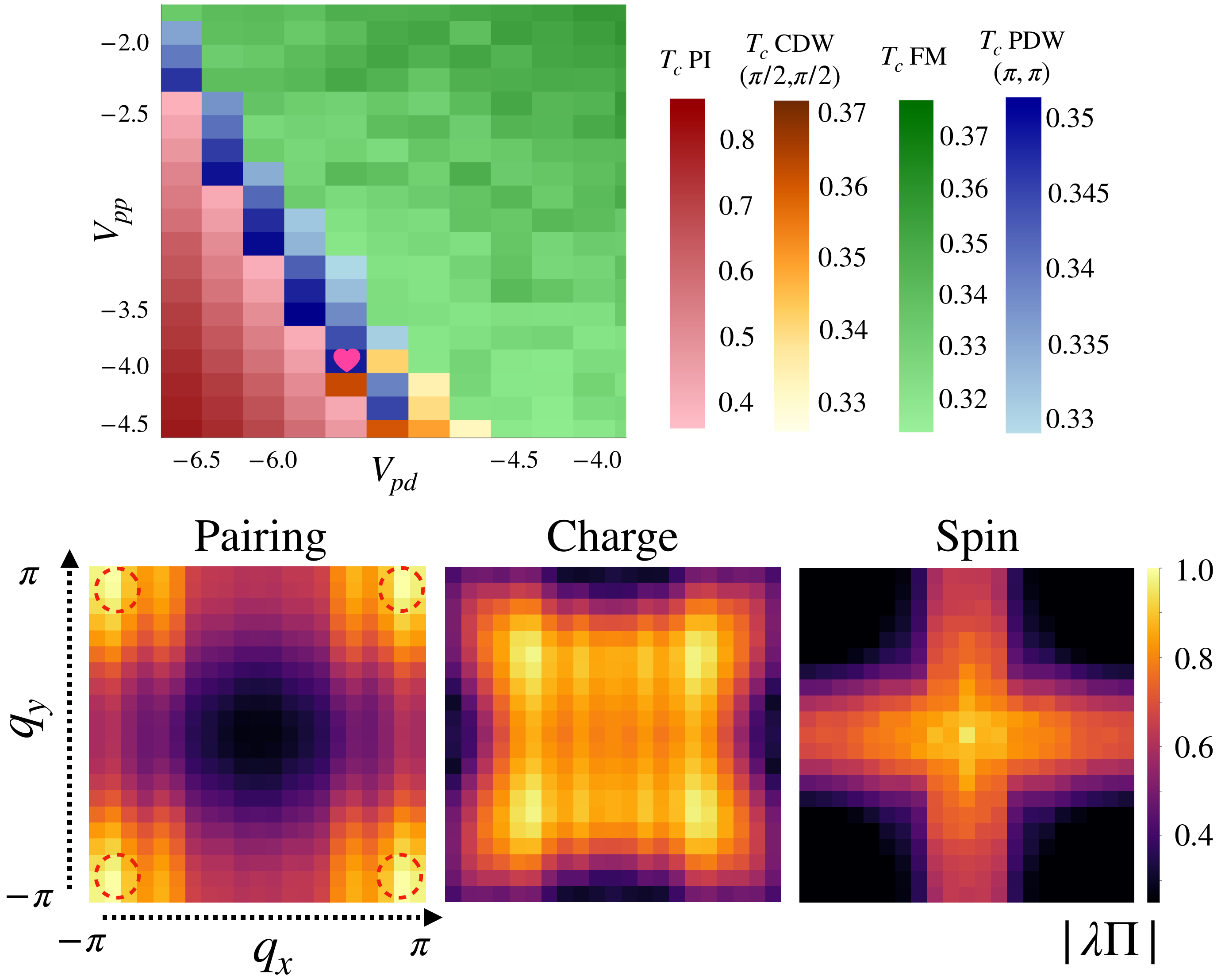}
\caption{(Top) Phase diagram for the parameters specified in the main text. Darker shading indicates a higher transition temperature. (Bottom) Coefficients to the pairing and charge order parameters as a function of ordering wavevector $\bq$ for $V_{pp}=-4,V_{pd}=-5.5,V_{pp}'=V_{dd}'=2$ at $T_c=0.34$. The location in the phase diagram corresponding to this choice of interaction strengths is marked by a pink heart. Wavevectors in the $(q_x,q_y)$ plane where the Stoner criterion is met are marked with red dashed lines. PDW = $(\pi,\pi)$ pair density wave, PI = Pomarenchuk instability, CDW = $(\pi/2,\pi/2)$ charge density wave, FM = ferromagnet.}
\label{fig:fig4}
\end{figure}
Analogously to the checkerboard model, we observe that the phase diagram of the three-band Hubbard model, shown in Fig.\ref{fig:fig4}, supports a PDW only when some form of nearest-neighbour attraction is present. These results are similar to those recently reported on DMRG studies of two-leg ladders \cite{HCthreeband}, with both attractive $V_{pd}$ and $V_{pp}$ found to be indispensable to PDW formation. However, as with the checkerboard model, we find that additional next-nearest-neighbour repulsive  interactions $V_{pp}', V_{dd}'>0$ are necessary for favouring the PDW over competing charge orders. We show the coefficients of the pairing and charge order parameters in Figure \ref{fig:fig4} for $V_{pp}=-4,V_{pd}=-5.5$, a choice of interaction strengths marked by a pink heart in the phase diagram. The pairing channel is the dominant of the three channels considered, with a peak near $\bq = (\pi,\pi)$. The pairing symmetry here is $s$-wave, just like the checkerboard model.

\section{\MakeUppercase{The role of quantum geometry}} 
\label{sec:the_role_of_quantum_geometry}
The quantum geometry of a band is characterized by its quantum geometric tensor (QGT) \cite{quantum_geometry},
\begin{equation}
    \begin{aligned}
        \mathcal{B}_{ij}(\bk) \equiv &\langle \partial_i u_{\bk}|\partial_j u_{\bk}\rangle - \langle \partial_i u_{\bk}|u_{\bk}\rangle\langle u_{\bk}|\partial_j u_{\bk}\rangle
    \end{aligned}
\end{equation}
The real part of this tensor, denoted $g_{ij}(\bk)$, is identified as the quantum (or Fubini-Study) metric, a positive semi-definite Riemannian metric that is used to distances in amplitude between proximate quantum states. The imaginary part of the QGT, meanwhile, corresponds to the Berry curvature, and is therefore related to differences in phase between quantum states. Various quantum transport and interaction phenomena can be related to the QGT, such as fractional quantum Hall effect \cite{parameswaran2013},  fractional Chern insulators \cite{neupert2015,liu2024}, and superfluidity in a flat band \cite{peotta2015, torma2022}, including for multiorbital superconductors \cite{PhysRevLett.131.016002,Kitamura2022,Chen2023}. In the latter case, it was shown how the total superfluid weight can become nonpositive definite due to pairing fluctuations in the presence of orbital-dependent interactions, thus facilitating the transition of a uniform BCS state to a PDW. 

Given the key role played by the momentum-dependent form factors $\{u_{ma}(\bk)\}$ in supplying COM dependence $\bq$ to the interaction, one might expect signatures of the PDW instability to appear in the QGT of the multiband system under consideration. In this section, we will compare the competing tendencies towards uniform and PDW superconducting order by presenting a set of conditions on the quantum metric $g$ that are sufficient---but not necessary---for favouring finite-momentum pairing interactions \footnote{Note that these conditions are not sufficient to prove the existence (or lack thereof) of a PDW \textit{instability}. In an $s$-wave flat band context, however, it can be shown that these conditions are indeed sufficient to yield a Stoner-type PDW instability because the susceptibility $\Pi$ is independent of $\bq$.}. 

To this end, assume that the dominant interaction is in the particle-particle channel, and as such the effective coupling $\lambda^\text{p}_{\bq}$ is attractive. Suppose we initially consider $\bq=(0,0)$ and perturb by a small amount $\delta\bq$. To second order, the interaction varies as \footnote{We know that first-order contributions must disappear because $V_{\bq}$ must, by symmetry considerations, possess a saddle-point at $\bq=0$}
\begin{equation}
   V^\text{p}_{\bk,\bp}(0)\to V^\text{p}_{\bk,\bp}(0)+\delta\bq_\mu\delta\bq_\nu\left(\left.\frac{\partial^2}{\partial_{\bq_\mu}\partial_{\bq_\nu}} V^\text{p}_{\bk,\bp}(\bq)\right)\right\rvert_{\bq=0} 
\end{equation}
In analogy to non-degenerate perturbation theory, the concavity of the most negative eigenvalue at $\bq=0$ can be approximated by
\begin{align}
    -\partial^2_{\bq_\mu\bq_\nu}|\lambda_{\bq}|_{\bq=0}  &= \sum_{\bk\bp} v^{*}_{0}({\bk})\left.\left(\partial^2_{\bq_\mu\bq_\nu}V^\text{p}_{\bk\bp}(\bq)\right)\right\rvert_{\bq=0}v_0(\bp),
\end{align}
where $v_0(\bk)$ is the eigenvector of $V^\text{p}_{\bk,\bp}(\bq=0)$ corresponding to the most negative eigenvalue. If $\partial^2_{\bq_\mu\bq_\nu}V_{\bk\bp}(\bq)$ is negative-definite, then the concavity of $\lambda$ at $\bq=0$ is also negative, and therefore a small variation in $\bq$ will result in a stronger effective coupling in the particle-particle channel \footnote{This is a sufficient but not necessary condition; it is, of course, possible that the concavity of $\lambda$ at $\bq=0$ is positive but that the global minimum of $\lambda$ still occurs at some nonzero $\bq$).}. This is favourable for PDW formation. 

If we make the simplifying assumption that the interaction carries no information about the orbital dependence, we may express the band-projected interaction in terms of overlaps between Bloch states:
\begin{equation}
    \begin{aligned}
        {V}^{\text{p}}_{\bk\bp}(\bq) &= V(\bk-\bp)\langle u_{\bk+\bq/2}|u_{\bp+\bq/2}\rangle\langle u_{-\bk+\bq/2}|u_{-\bp+\bq/2}\rangle
    \end{aligned}
\end{equation}
To make analytical progress, assume that $V(\bk-\bp)$ is sharply peaked around $\bk\approx \bp$. This approximation is valid in the limit of long-range interactions. For example, suppose that $V(\bk-\bp)$ can be modelled by a simple two-dimensional Gaussian distribution centred at zero,
\begin{equation}
    \begin{aligned}
        V(\bk-\bp) = -\frac{|V_0|}{2\pi}&\det (\vec{\Sigma})^{-1/2}e^{-\frac{1}{2}(\bk-\bp)^T\vec{\Sigma}^{-1}(\bk-\bp) },
    \end{aligned}
\end{equation}
where $\vec{\Sigma}_{ij}=\xi_0^2$ is a rotationally-isotropic covariance matrix. The sign of $V$ is fixed to be attractive, reflecting the fact that we are in the superconducting phase. Because only $\bk\approx \bp$ contributions are picked out by the interaction, we may express the Bloch state overlaps in terms of the quantum metric \footnote{Note that the quantum metric depends not only on the tight-binding parameters of the lattice, but also on the locations of the orbitals \cite{huhtinen2022} within the unit cell. To ensure that the quantum metric is uniquely determined, here we fix the orbitals to be located at their physical intra-unit-cell positions, with the density-density interaction depending only on the actual distances between orbitals.}. The concavity of the most negative eigenvalue is given to leading order by
\begin{equation}
        - \partial^2_{\bq_\mu\bq_\nu}|\lambda_{\bq}|_{\bq=0}\propto |V_0|\xi_0^4 \sum_{\bk} |v_0(\bk)|^2\partial^2_{\bk_\mu\bk_\nu} \Tr[G(\bk)],
\end{equation}
where $G_{ij}(\bk)\equiv \left[ g_{ij}\left(\bk\right)+g_{ij}\left(-\bk\right)\right]/2$. (Under time-reversal or inversion symmetry, then $G_{ij}(\bk)=g_{ij}(\bk)$). We can thereby conclude that the concavity of $\lambda$ is determined by the local concavity of the quantum metric. Specifically, for attractive long-range interactions, we find that negative concavity in the quantum metric is favourable for PDW formation. This result joins a growing body of work suggesting the relevance of quantum geometry for promoting various density wave orders, such as in multiorbital systems with orbital-dependent pairing \cite{PhysRevLett.131.016002} and in flat band systems with quantum geometric nesting \cite{han2024}. Combining these schemes, one can envision engineering Hamiltonians with quantum geometries optimally tuned for supporting PDW states. 

\section{\MakeUppercase{Conclusions}}
\label{sec:conclusion}
In this work we have outlined a general mean field scheme for eliciting PDW order from multiband physics beyond weak coupling. The onset of PDW superconductivity requires that i) the attractive pairing interaction persists at finite COM momentum, $\bq$, and ii) the interaction is strong enough to trigger the Stoner-type instability. We have shown that for a system dominated by density-density interactions, the first condition can be met in multiband systems where the band-projected interaction naturally inherits COM $\bq$-dependence through the form factors. Using the checkerboard lattice and a three-band extended Hubbard model, we have explicitly revealed how intraband PDW orders can dominate over uniform superconductivity and other particle-hole condensates when nearest-neighbour attraction and next-nearest-neighbour repulsion are present. Our mean field results are consistent with recent DMRG works on multiband systems \cite{jiang2023pair,HCthreeband}, in which nearest-neighbour attraction was also found to be essential to the stability of the PDW state. The interactions considered here are also reminiscent of those presented in \cite{PhysRevLett.130.026001}, where it was found that spatially non-monotonic BCS interactions are indispensable to PDW formation. This work highlights the joint importance of multiband physics and nearest-neighbour interactions in inducing an attractive pairing interaction at finite Cooper pair momentum, thereby stabilizing the resulting PDW. 

\begin{acknowledgments}
    We thank Zhaoyu Han, Julian May-Mann, Trithep Devakul, John Sous, and Hong-Chen Jiang for insightful discussions.
    N.S.T. and S.R. are supported in part by the US Department of Energy, Office of Basic Energy Sciences, Division of Materials Sciences and Engineering, under Contract No. DE-AC02-76SF00515. Y.-M.W. acknowledges the Gordon and Betty Moore Foundation’s EPiQS Initiative through GBMF8686 for support.
\end{acknowledgments}

\bibliography{pdwmultiband}

\end{document}